\documentclass[5p]{elsarticle}

\usepackage[T1]{fontenc}
\usepackage{epsf,graphicx}
\usepackage{natbib}
\usepackage{multirow}
\usepackage{amsmath,gensymb,amssymb}
\usepackage{siunitx}
\usepackage[bottom]{footmisc}
\usepackage{lipsum}
\usepackage{graphicx}
\usepackage{dcolumn}
\usepackage{upgreek}

\usepackage{bm}
\usepackage{flushend}
\usepackage{color}
\usepackage{xcolor}
\usepackage{esvect}
\usepackage{physics}
\usepackage{tcolorbox}
\usepackage[shortlabels]{enumitem}
\usepackage[colorinlistoftodos]{todonotes}

\usepackage{hyperref}
\hypersetup{
colorlinks=true, 
citecolor=magenta,
breaklinks=true, 
urlcolor= blue, 
linkcolor= blue, 
bookmarksopen=true, 
}

\newcommand{\eff}{_{\mathrm{eff}}}

\journal{Physica D: Nonlinear Phenomena}

\begin{document}

\begin{frontmatter}

\title{Nonlinear dynamics of a hanging string with a freely pivoting attached mass}

\author[1]{Filip Novkoski}
\author[1,2]{Jules Fillette}
\author[3]{Chi-Tuong Pham}
\author[1]{{Eric Falcon}\corref{cor1}}
\ead{eric.falcon@u-paris.fr}

\affiliation[1]{organization={Universit\'e Paris Cit\'e, CNRS, MSC, UMR 7057}, city={Paris}, postcode={F-75013}, country={France}}
\affiliation[2]{organization={LPENS, ENS, CNRS, UMR 8023, PSL Research University, Sorbonne Université}, postcode={F-75005},city={Paris},country={France}}
\affiliation[3]{organization={Universit\'e Paris-Saclay, CNRS, LISN, UMR 9015}, city={Orsay}, postcode={F-91405}, country={France}}

\cortext[cor1]{Corresponding author}

\begin{abstract}We show that the natural resonant frequency of a suspended flexible string is
  significantly modified (by one order of magnitude) by adding a freely pivoting attached mass at
  its lower end. This articulated system then exhibits complex nonlinear dynamics such as bending
  oscillations, similar to those of a swing becoming slack, thereby strongly modifying the system
  resonance that is found to be controlled by the length of the pivoting mass. The dynamics is
  experimentally studied using a remote and noninvasive magnetic parametric forcing. To do so, a
  permanent magnet is suspended by a flexible string above a vertically oscillating conductive
  plate. Harmonic and period-doubling instabilities are experimentally reported and are modeled
  using the Hill equation, leading to analytical solutions that accurately describe the
  experimentally observed tonguelike instability curves.
\end{abstract}




\end{frontmatter}



\section{Introduction}
The vibrational modes of a flexible and inextensible hanging string in the gravitational field is a
classical mechanical problem, and its solution dates back to Bernoulli and Euler in the 18th century
\cite{Bernoulli1738}. The case of the flexible string with an additional mass at its lower end has
been considered only recently, both theoretically~\cite{ArmstrongAJP1976} and
experimentally~\cite{Deschaine2008}, and its modes are found to be close to those of a rigid
pendulum~\cite{Deschaine2008}. {Although such a mechanical system involves a spatially dependent
  wave velocity (as for acoustics in air ducts~\cite{ColtmanJASA1979} or in granular
  media~\cite{LiuPRL1992}), there is little experimental data for such articulated pendulum systems
  compared to those involving rigid limbs such as the double pendulum~\cite{Lee1970}.}

Here, we {study a system constituted of a flexible string with a freely pivoting attached mass to
  its lower end. We show that its natural resonant frequency is one order of magnitude higher than
  the one of a single flexible string. Such a two-component articulated system exhibits} bending
oscillations similar to those of a swing becoming slack, thus strongly modifying its
resonances. They are studied using {an original technique based on} noninvasive and remote magnetic
forcing. More precisely, we investigate the dynamics of a permanent magnet{, freely pivoting,}
suspended by a flexible string above a vertically oscillating conductive plate. Induced currents due
to the moving conductor generate remote forces on the magnet, giving rise to a parametric forcing
{needed to study the nonlinear dynamics of the system}. {The main advantage of this forcing is to
  avoid the excitation of the string torsional modes and the single pendulum mode unlike vertically
  oscillating the top attachment point of the string. Note that} parametric
instabilities~\cite{Boeck2020}, phase transitions~\cite{Schmidt1984}, bifurcations~\cite{Kim1997}
and control of chaos~\cite{Starrett1974} have been investigated with a {\em rigid} pendulum with a
magnet as their tip mass {in an oscillating magnetic field}~\cite{Luo2020,Khomeriki2016}, or with a
ferrofluid drop suspended by a torsion pendulum~\cite{Shliomis2004}. For a {\em flexible} string,
{\em with no tip mass} and vertically vibrated, parametric resonance, chaos, and self-knotting have
been experimentally evidenced~\cite{Belmonte2001}.

\begin{figure}[t!]
  \includegraphics[width=\columnwidth]{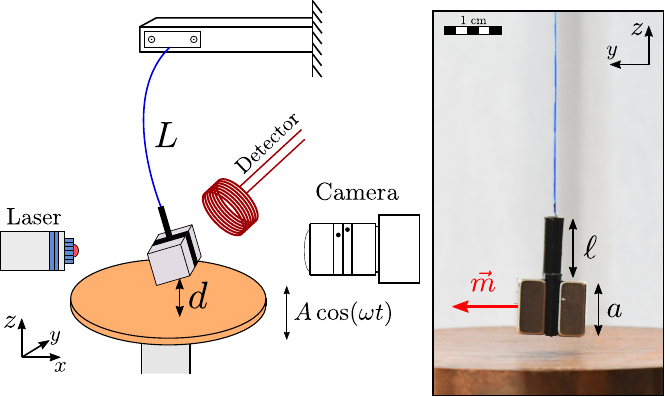}
  \caption{\label{fig:setup}Left: Illustration of the experimental setup with the system in motion. Right: Photo of the 
    magnets suspended at the end of a nylon cord, using the plastic holder of length $\ell + a$. $L=6$ cm.}
\end{figure}

\section{Experimental setup}
The setup is shown in Fig.~\ref{fig:setup}. It consists of a {flexible} nylon cord suspended at its
upper point and with a tip mass attached at its lower end using glue. The cord has a 0.35-mm
diameter, linear mass density $\rho=1.24$ mg/cm, and length $L=6$ cm. The tip mass consists of two
neodymium permanent magnets (square cuboid of size $a\times a\times \frac{a}{2}$, with $a=1$ cm,
total magnetic moment $\mathrm{m}=0.8\pm0.05$ A/m$^{2}$) in which a 3D-printed thin plastic square
plate is sandwiched with a length of $\ell\in[0,2]$ cm protruding above the magnets (see
Fig.~\ref{fig:setup}). The total tip mass is $M=8$ g.  {The articulated system is thus constituted
  of a flexible string (of length $L$) with a rigid pendulum (of length $\ell+a$) attached to its
  end. We force the system parametrically, using a noninvasive and remote magnetic forcing, to study
  its dynamics. To do so, the system is suspended above} the center of a circular copper plate. The
plate is attached to an electromagnetic shaker using a long aluminum shaft to avoid any
interference from the internal magnetic field of the shaker. The plate is subjected to vertical
oscillations, $A\sin(2\pi f t)$, at a mean distance $d_0\in[7,9]$ mm from the magnet center, with
$f\in[15,40]$ Hz and $A\in[0.07,2]$ mm, the frequency and amplitude of the vibrating plate{,
  respectively}. The relative distance between the magnets and the plate is
$d(t)=d_0+A\sin(2\pi f t)$. The horizontal velocity of the magnet is measured either by a Polytech laser vibrometer
for small amplitude motions or by a homemade copper wire coil, placed close to the magnet, for large
ones. The magnet motion is visualized by a Basler camera ($2048\times1536$ px$^2$, 120 fps) located
in front of it.

\begin{figure}[t!]
  \includegraphics[width=\columnwidth]{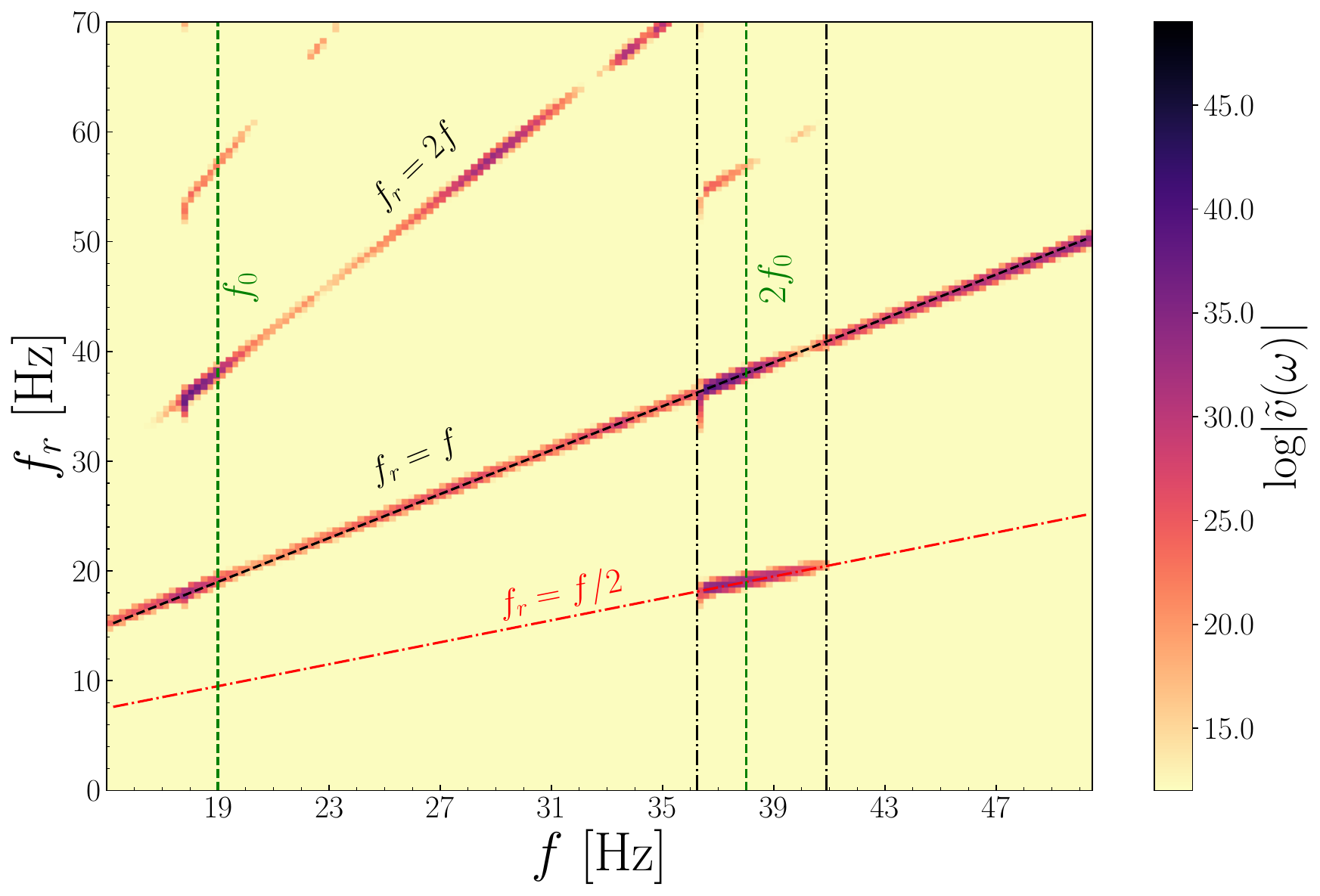}
  \caption{\label{fig:spectrogram}Spectrum of the magnet velocity showing the system frequency response, $f_r$, for different forcing frequencies, $f$. $f_0$ is the system's natural resonant frequency. $\ell=6$ mm. A subharmonic response, i.e., $f_r=f/2$, occurs for frequencies around $2f_0$ (between the dash-dotted lines). Note also the presence of the second harmonic (i.e., $f_r=2f$) that is maximal around $f_0$.}
 \end{figure}

 \section{Frequency response}
 We first determine the frequency response of the magnet-string system. To do so, the forcing frequency $f$ is varied
 linearly in time, during $2$ min, from $15$ to $50$ Hz, and the horizontal velocity $v(t)$ of the
 tip mass is recorded. We compute its frequency spectrum
 $S_v(\omega)\sim |\hat{v}(\omega)|$ [where $\hat{v}(\omega)$ is the Fourier transform of $v(t)$]
 as shown in Fig.~\ref{fig:spectrogram}. The spectrum maxima correspond to the system frequencies
 $f_r$ in response to different forcing frequencies $f$. We clearly see that the response follows
 the forcing, $f_r=f$, except in a region (see vertical dash-dotted lines), where a subharmonic
 behavior, $f_r=f/2$, is observed. In this range, the magnet exhibits large-amplitude oscillations
 (in the planar pendulum motion) in which the string swings back and forth at half the driving
 frequency, corresponding thus to a period-doubling instability [see the red region and photos in Fig.~\ref{fig:seuil}, and a movie in the Supplemental Material (see~\ref{AppendixA})]. The system
 exhibits also small rotations of the {\em center of mass}, around the equilibrium, with deviations
 of the string from the vertical. Such motion is restricted to the vertical $y-z$ plane and no
 rotation around the vertical axis is found.  Outside the red region in Fig.~\ref{fig:seuil}, a low-amplitude motion is
 observed. Interestingly, the magnetic damping of the conductive plate cancels any torsional
 motion. 
 
 We now measure the natural resonant frequency $f_0$ of the magnet-string system. We move laterally
 the magnet away (over the $y$-axis) from its equilibrium with a small string flexion to excite only
 the mode observed in the inset of Fig.~\ref{fig:seuil} and not the unflexible pendulum one
 [occurring at a much lower frequency $\sqrt{g/L}/(2\pi) \simeq 2$ Hz]. We record the magnet
 velocity during its free-damped oscillations. Through a Fourier transform, the resonant frequency
 is inferred from the spectrum peak frequency, $f_0$, and is noted by a vertical dashed line in
 Fig.~\ref{fig:spectrogram}. The subharmonic region in Fig.~\ref{fig:spectrogram} occurs around
 $2f_0$, as expected for parametric resonance~\cite{Landau1960}, with the corresponding response
 $f_r=f/2$, i.e., a period-doubling motion. Figure~\ref{fig:spectrogram} also shows that no
 subharmonic instability occurs around $f_0$, as expected, whereas a second-harmonic instability,
 i.e., $f_r=2f$, is observed.

\begin{figure}[t!]
  \includegraphics[width=\columnwidth]{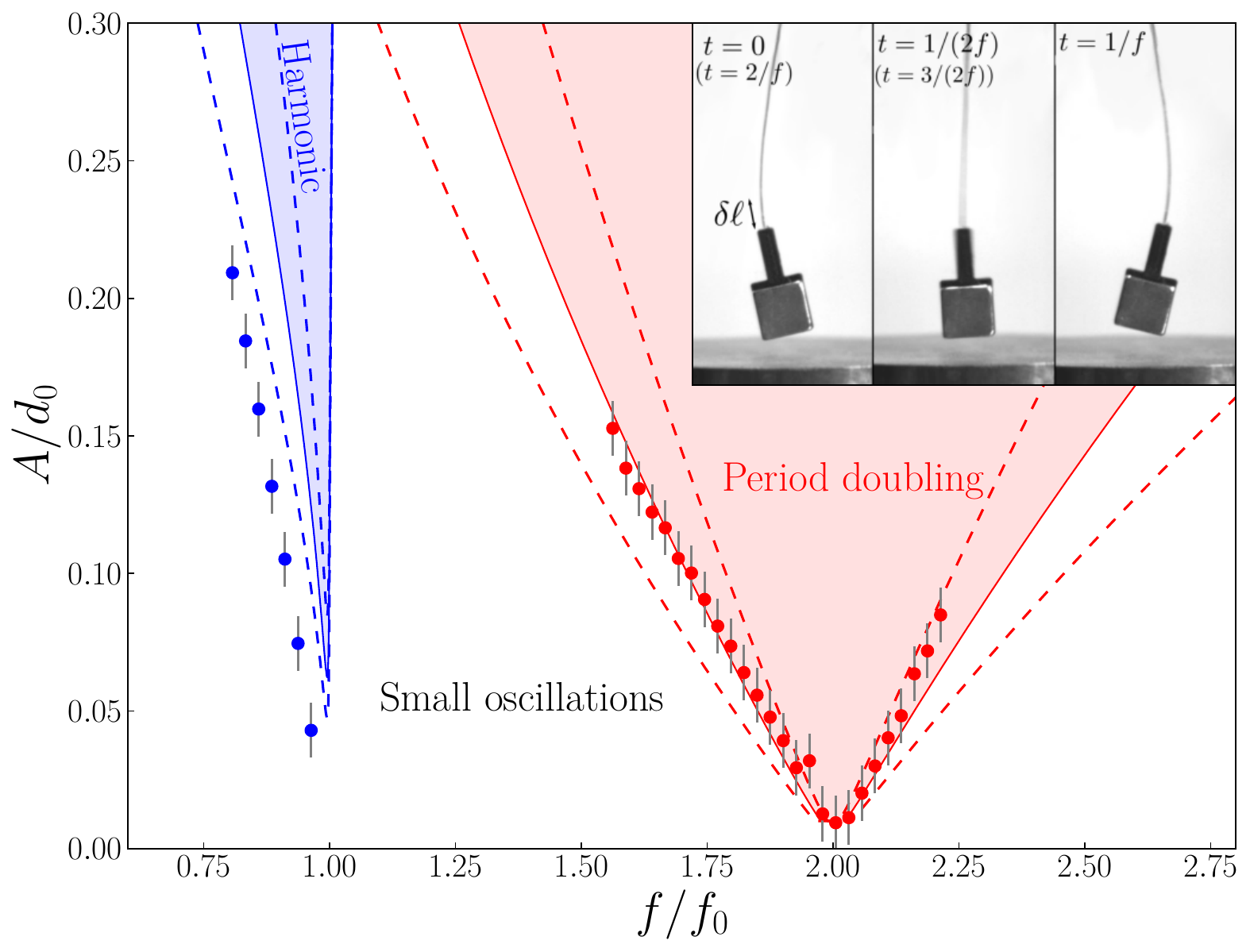}
  \caption{\label{fig:seuil}Dynamic behaviors {of the articulated system} in the nondimensional forcing amplitude and frequency
    parameter space. Two resonant tongues around $f_0$ and $2f_0$ are observed, corresponding to a
    harmonic instability and period doubling, respectively. Bullets: experimental data. Solid lines: prediction of the resonant tongues
    from Eq.~\eqref{eq:tongue} (red) and Eq.~\eqref{eq:solharmo} (blue). Dashed lines: errors in the predictions due to the estimation of
    $q$. Inset: Photos at different times showing period-doubling instability and a strong
    deviation angle of the magnet and correspondingly of the string. $\ell=1$ cm. $L=6$ cm.}

 \end{figure}
 
\section{Instability threshold}
The threshold of the subharmonic instability depends on the vibrating parameters $f$ and $A$. It is
experimentally measured by slowly increasing $A$, for a given $f$, until observing an exponential
rise in the system oscillation amplitude. The experimental data are shown in Fig.~\ref{fig:seuil} to
collapse onto continuous curves in the nondimensional parameter space ($A/d_0$, $f/f_0$) which are
suggestive of two resonant curves around $f_0$ (in blue) and $2f_0$ (in red). In the zone above the
red points, a period doubling of the oscillations of the bending string is observed, i.e.,
$f_r=f/2$, corresponding to the inset of Fig.~\ref{fig:seuil}.  The tonguelike shape of this
marginality curve is reminiscent of parametric resonance. The instability is also found slightly
hysteretic. In the region above the blue points, the system responds, as $f_r=f$, corresponding to a
harmonic instability.  Note that, for frequencies slightly close and above $f_0$, no experimental
data are indicated since we do not observe a clear transition to the harmonic instability (see
below).

\section{Bending string description}\label{string}
The behavior of a hanging flexible string
 differs significantly from that of a rigid pendulum~\cite{Bernoulli1738}. Let us consider a
 flexible and inextensible string suspended from its fixed upper point, subjected to gravity and
 tension forces, and a maximal displacement of its end $X_0$ at an initial time. The string position
 $\mathbf{X}(s,t)$ is then described by Newton's law as~\cite{Belmonte2001}
\begin{align}\label{eq:pde}
  \rho\partial_{tt}\mathbf{X}=\partial_s[T(s,t)\mathbf{\hat{s}}]-\rho g\mathbf{\hat{z}}
\end{align}
where $s\in(0,L)$ is the curvilinear position along the string taking its origin at its upper end
[$\mathbf{X}(0,t)=\mathbf{0}$], $T(s)$ is the tension acting in the string, $\rho$ is the linear
mass density, and $g=9.81$ m/s$^2$. By applying additional boundary conditions (BCs), $T(L,t)=0$ (no
tension acts on the free end), and $\mathbf{X}(L,0)=\mathbf{X_0}$ (initial condition), the solutions
of Eq.~\eqref{eq:pde}, for no tip mass and small displacements from the equilibrium configuration,
are found as modes of the zeroth-order Bessel function of the first kind
$J_0$~\cite{Bernoulli1738}. Nevertheless, due to the added tip mass, our system is able to rotate
around its {\em center of mass} (in the planar pendulum motion), and a significant modification of
its resonant frequencies is predicted~\cite{Deschaine2008}. Beyond BCs on the string position at its
ends, forces, and torques on the tip mass need to be theoretically considered leading to nontrivial
BCs~\cite{Deschaine2008}. In that case, solutions of Eq.~\eqref{eq:pde}, for small displacements,
are that of a classical harmonic oscillator of frequencies $\omega_n$, each corresponding to a
system mode. Note that if the string were elastic, one would recover a parametrically
  excited buckling beam, most likely governed by a Duffing-like equation~\cite{Abou1992}.

By considering small motions around equilibrium, Eq.~\eqref{eq:pde} reads
\begin{align}
  \rho\partial_{tt}X=\partial_z\left[\left(\rho gz + Mg\right) \partial_zX \right],
\end{align}
where $X$ is the $x$-direction component of $\mathbf{X}$. Through separation of variables $X=p(t)q(z)$
and assuming that the hanging mass is much larger than the mass of the cord, one finds
\begin{align}
  \partial_{tt}p=-\omega^2_np.
\end{align}
The boundary conditions imposed on the cord determine exactly the value of $\omega_n$. If
we assume now that the hanging mass experiences a time-dependent form, and that the gravitational
constant is thus given as $g(t)=g[1+a(t)]$, we find through the same procedure as above
\begin{align}\label{eq:final}
  \partial_{tt}p=-\left[1+a(t)\right]\omega^2_np,
\end{align}
where $a=F(t)/(Mg)$, and $F$ is the unknown vertical force exerted on the hanging mass $M$. An
important assumption was made above, namely that a separation of time and space variables is possible, which is no
longer the case for large tip-mass displacements. Additionally, Eq.~\eqref{eq:final} is a significant
simplification considering the nontrivial BCs at the lower end~\cite{Deschaine2008}.
Ideally, one would have to also include the remote force exerted by the plate on the magnet into the bottom
boundary condition, notably in the acceleration of the center of mass.

\begin{figure}[t!]
  \includegraphics[width=\columnwidth]{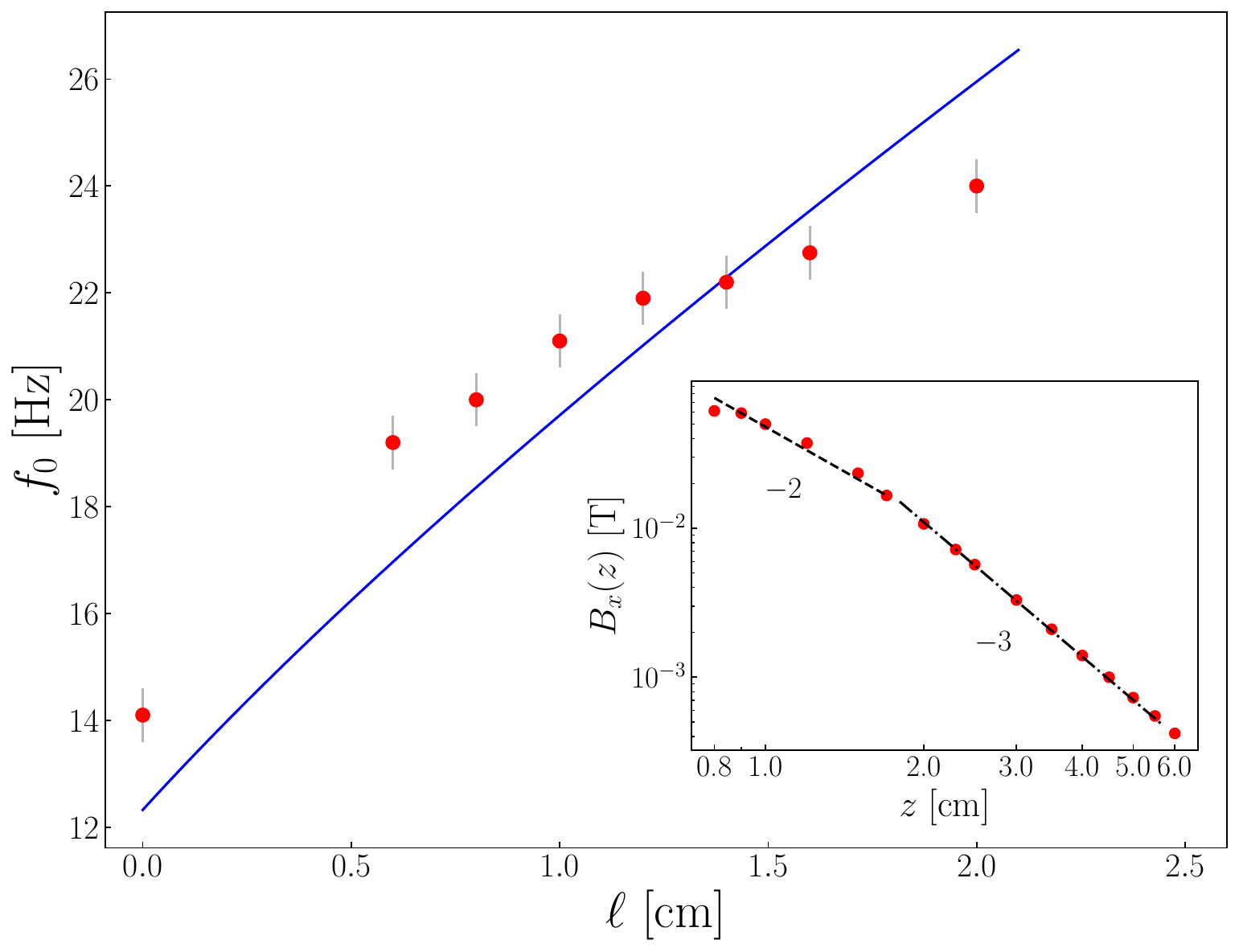}
  \caption{\label{fig:resonance}Experimental resonant frequencies, $f_0$, of the tip-mass-cord system
    for different {pivoting mass} lengths $\ell$. The solid line corresponds to Eq.~\eqref{eq:pde} of the flexible string
    model with a tip mass, i.e., including nontrivial BCs. Inset: Magnetic field $B_x$ measured at different distances $z$ below the
    magnet bottom without forcing (bullets). Near-field dipole prediction (dashed line) and far-field monopole
    model (dash-dotted line). $L=6$ cm.}
 \end{figure}

\section{Pivoting-mass length effects}
We now measure experimentally the resonant frequencies
$f_0$ for different lengths $\ell$ as shown in Fig.~\ref{fig:resonance}. A strong dependence of
$f_0$ on $\ell$ is found, and $f_0$ is significantly larger (by one order of magnitude) than the
corresponding usual rigid pendulum mode ($2$ Hz) or the no-tip-mass flexible string mode ($2.4$
Hz). The numerical computation of $f_0$ using Eq.~\eqref{eq:pde} and the BCs found in
Ref.~\cite{Deschaine2008} is shown as well, which roughly agrees with the data with no fitting
parameter. We thus find that the tip-mass flexible string is reasonably well modeled by Eq.~\eqref{eq:pde} with the BCs 
in Ref.~\cite{Deschaine2008}.  As
the string is not able experimentally to freely rotate around its lower attachment point (see inset
of Fig.~\ref{fig:seuil}), a rigid motion occurs on the string $\delta\ell\simeq2$ mm away from its
lower end, and we have thus introduced an effective length of the axis, $\ell\eff=\ell+\delta\ell$,
in this model.

\section{Hill's equation}
Let us now take into account the remote magnetic forcing as the vertical oscillations of the
conducting plate induce a time-dependent force on the magnet. First, we measure, with no forcing and
using a Koshava5 Gaussmeter, the field amplitude $B_x(z)$ generated by the magnet along the negative
$z$ axis (i.e., below the magnet and in the direction of the magnetic dipole moment
$\mathbf{m}$). Far from the magnet, $B_x(z)$ is well described (see inset of
Fig.~\ref{fig:resonance}) by the classical magnetic dipole law
$B_x(z)=\mu_0 \mathrm{m}/(4\pi z^3)$~\cite{Jackson}, with $\mu_0=4\pi$10$^{-7}$ Tm/A the vacuum
magnetic permeability. By moving closer to the magnet, the experimental magnetic strength clearly
departs from this prediction and is roughly fitted by $B_x(z)=\mu_0q/(4\pi z^2)$ where $q$ is an
effective monopole moment inferred from the fit as $q=33\pm 4$~Am. Note that the components $B_y$
and $B_z$ are negligible due to the choice of $\mathbf{m}$ along the $x$ direction, since they are measured to be ten times
  smaller than $B_x$. Additionally, more sophisticated expressions for the field of a permanent magnet could be
  used~\cite{Furlani001}, but for the sake of simplicity we rely on this empirical model,
  also known as the dumbbell model~\cite{Concha2018}. The electromagnetic interaction between the
magnetic field $B_x$ and the vertical motion of the nearby oscillating conductor causes a vertical
force on the magnet computed as $F_z(d) = (\mu_0^2q^2\sigma u_z)/(16\pi d)$ (see~\ref{AppendixB})
with $d(t)=d_0+A\sin{(2\pi ft)}$ the relative distance between the magnet and the plate,
$u_z=\mathrm{d}d/\mathrm{d}t$ their relative vertical velocity, and $\sigma$ the plate
conductivity. By adding the driving force $F_z$ into Eq.~\eqref{eq:pde} of the hanging string,
{including the nontrivial BCs}, and assuming small horizontal displacements, we find that the
amplitude $p_n$ of each mode $n$, oscillating at $\omega_n$, is governed by an equation of the Hill
type (see \ref{AppendixD})
\begin{align}\label{eq:hill}
  \partial_{tt}p_n=-\omega^2_n\left[1+c\frac{(A/d_0)\omega\cos(\omega t)}{1+(A/d_0)\sin(\omega t)}\right]p-\kappa\partial_tp_n
\end{align}
with $c=(\mu^2_0q^2\sigma)/(16\pi Mg)=0.026$ s, and $\kappa$ a dissipation coefficient. It is
well-known that the Hill equation can lead to resonant tongues of parametric instability including the
special case of the Mathieu equation~\cite{Chicone2006}.

\begin{figure}[t!]
  \includegraphics[width=\columnwidth]{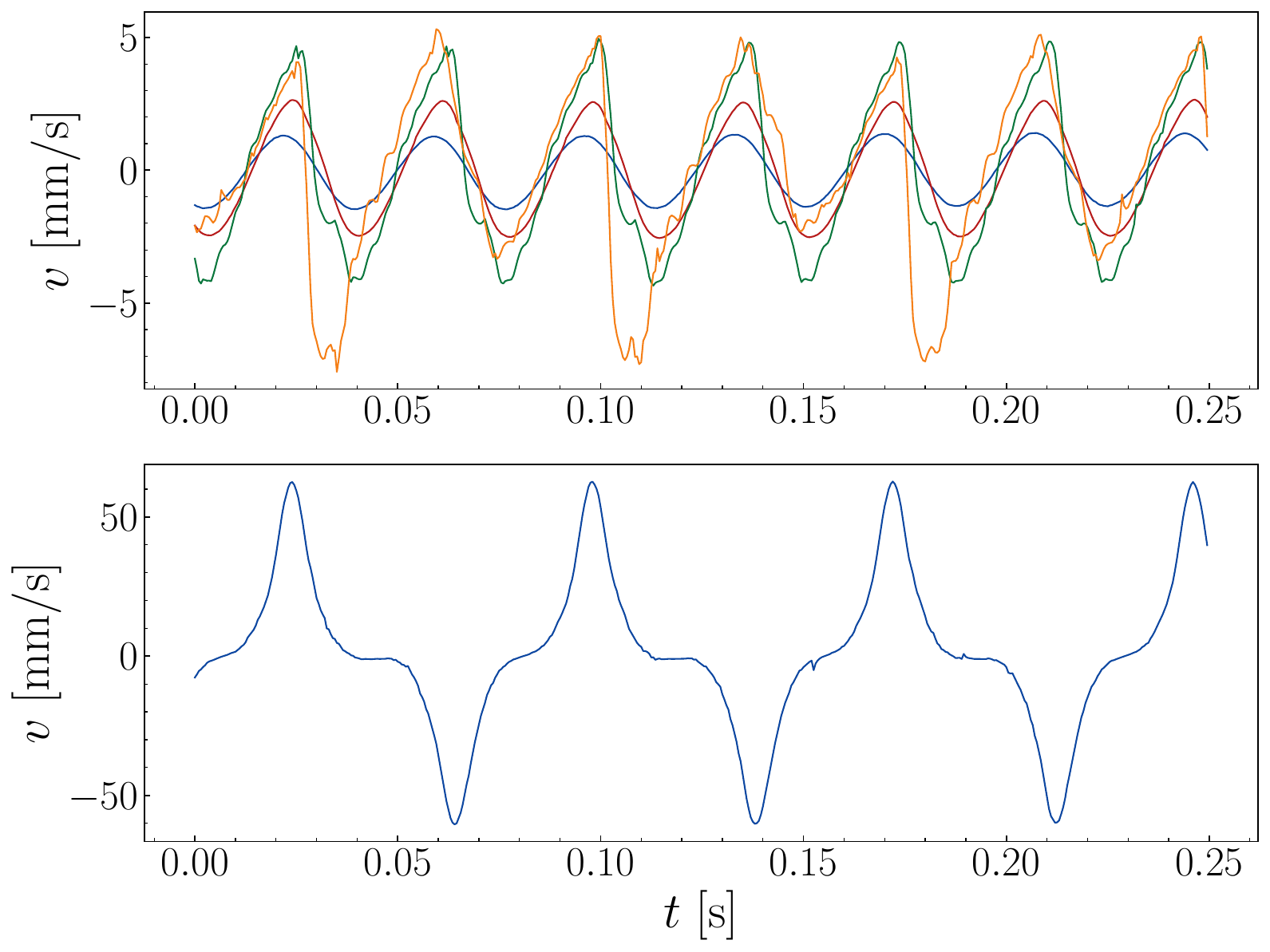}
  \caption{\label{fig:period}Top: Temporal evolution of the horizontal velocity of the magnet, for
    increasing plate amplitudes $A$ (in order blue, red, green, yellow) at fixed $f=27$ Hz, and
    below the period-doubling instability threshold (white area in Fig.~\ref{fig:seuil}). Bottom:
    same within the period-doubling region (red area in Fig.~\ref{fig:seuil}).}
\end{figure}

By applying a Fourier expansion of the periodic function {within the brackets in}
Eq.~\eqref{eq:hill} up to the second order, we can analytically find the expressions for the
resonant tongues in the ($f/f_0$, $A/d_0$) parameter space (see~\ref{AppendixD}). The
tongues originate at $2\omega_n/m$ ($m=1$, 2,~... is the tongue number) leading thus to a harmonic
or subharmonic response. We reproduce here only the expression for the subharmonic tongue at $2f_0$, i.e.,
\begin{align}\label{eq:tongue}
   \left(\frac{2\omega^2_0}{\omega^2}-\frac{1}{2}\right)^2+\frac{\kappa^2}{\omega^2}-\left(c e_1\frac{\omega_n^2}{\omega}\right)^2=0.
\end{align}
where $e_1$ corresponds to the first Fourier coefficient in the series expansion, which depends
nontrivially on the forcing amplitude (the exact expressions being cumbersome, they have been collected in~\ref{AppendixD} and~\ref{AppendixE}). The subharmonic instability curve of Eq.~\eqref{eq:tongue} is
plotted in Fig.~\ref{fig:seuil} (red solid line) along with the harmonic prediction at $f_0$ (blue
one) from Eq.~\eqref{eq:solharmo}. Our model is found to be in good agreement with the experimental instability curves with no
fitting parameter once the tongue minimum is known. As the latter is related to
dissipation~\cite{Verhulst2022}, this leads to an estimation of the effective dissipation $\kappa=0.18$
s$^{-1}$ satisfying the timescale separation $1/f \ll 1/\kappa$. Slight deviations of the model for higher frequencies (compare red solid line and bullets) could be due to the mode truncation to obtain the analytic expression. In addition, the theoretical tongue shape near $f_0$ (blue solid line) exhibits a sharp, almost vertical, rise above $f_0$, explaining why we are unable to experimentally observe the
right part of the harmonic instability curve.







\section{Transition behavior}
We finally focus on the magnet dynamics when crossing the subharmonics instability threshold. For this, we increase the plate amplitude $A$ at a fixed frequency $f$, and, using the laser vibrometer, we measure the horizontal velocity $v(t)$ of the magnet-string
  motion, as shown in top Fig.~\ref{fig:period}. For low $A$, small-scale oscillations are reported
  (see blue curve). For high enough $A$, still below the subharmonic instability threshold $A_c(f)$ in
  Fig.~\ref{fig:seuil}, we observe the appearance of higher harmonics as well as an asymmetry of
  the motion on every second period of the forcing frequency. This asymmetry is likely because the magnet velocity is no longer negligible. Once we cross the threshold of instability, the motion becomes more regular and with an
  amplitude that is ten times larger, as seen in bottom~Fig.~\ref{fig:period}, with period doubling
  clearly visible.  Interestingly, the magnet spends a long time at its extreme points, i.e., for $v=0$, with
  a rapid pass through its equilibrium point. Thus, increasing the amplitude means that higher harmonics grow
  more and more until the system transitions into the period-doubling state. This
  behavior is similar to that of the Duffing equation and indicates the presence of nonlinearities
  in the string.

\section{Conclusion}
We have reported, both experimentally and theoretically, the dynamics of an articulated oscillating
  system consisting of a flexible string with a freely pivoting mass at its end. The system resonant
frequency is controlled by the length of the pivoting mass and differs significantly (by one order
of magnitude) from that of the single flexible string. This articulated system in
  particular  exhibits bending oscillations, similar to a swing becoming slack. To do so, we used a
  remote and noninvasive parametric forcing technique. A magnet suspended by a flexible string
above a vertically oscillating conductive plate is subjected to a remote force (due to induced
  currents in response to the conductor motion). Parametric resonances of the tip-mass-string
system are then highlighted such as harmonic and period-doubling instabilities. This parametric system is theoretically modeled using a Hill equation and the obtained analytic expressions for the resonant tongues describe well the observed instabilities. Beyond their
thresholds, more complex dynamics can be observed once the magnet displacement or velocity becomes
comparable to the oscillating plate
one. A potential improvement to the current study would be to consider different magnet strengths. 






\section*{Acknowledgments}
We thank {V. Leroy for discussions and} A. Di Palma, Y. Le Goas, and O.-S. Souramasing for technical help. This work was supported by the Simons Foundation MPS No.~651463--Wave Turbulence (USA) and the French National Research Agency (ANR Sogood Project No.~ANR-21-CE30-0061-04 and ANR Lascaturb Project No.~ANR-23-CE30-0043-02).

\appendix

\section{Movie}\label{AppendixA}
 Video of the string, swinging back and forth at half the driving frequency, and showing a period-doubling instability. $f=30$ Hz.
 \href{https://ars.els-cdn.com/content/image/1-s2.0-S0167278924001155-mmc1.mp4}{See the video}

\section{Force on the magnet}\label{AppendixB}
In our model, we consider that the magnet is the source of a simplified magnetic field close to the
conducting plate in the form of
\begin{align}\label{eq:field}
  \vec{B}=\frac{\mu_0}{4\pi}\frac{q}{r^2}\hat{x},
\end{align}
which we find to be a reasonable approximation empirically. Following Ref.~\cite{Boeck2020}, the force on
the source of this field can be obtained through
\begin{align}\label{eq:force2}
  \vec{F} = \int_{\Sigma} \sigma(\vec{u}\times\vec{B})\times\vec{B} \textup{d}V,
\end{align}
where $\Sigma$ is the conductor volume, considered to be located at a distance $d$ from the source
and filling the half-infinite volume (we find numerically that effects of finite size can be
neglected due to the $1/r^2$ drop-off of field strength).

Inserting Eq.~\eqref{eq:field} into Eq.~\eqref{eq:force2} and considering $\vec{u}=u\hat{z}$, Eq.~\eqref{eq:force2} reads
\begin{align}
  \vec{F} = \left(\frac{\mu_0q}{4\pi}^2\right)\sigma u \int_{\Sigma} \frac{(\hat{z}\times\hat{y})\times\hat{y}}{r^4}\, \textup{d}V.
\end{align}
By turning to cylindrical coordinates, we have
\begin{align}
  \vec{F} = \left(\frac{\mu_0q}{4\pi}\right)^2\sigma u\hat{z} \int_d^\infty\int_0^\infty\int^{2\pi}_0 \frac{\rho}{\rho^2+z^2}\, \textup{d}\phi\textup{d}\rho\textup{d}z,
\end{align}
finally yielding
\begin{align}\label{eq:force}
  \vec{F} = \left(\frac{\mu_0q}{4}\right)^2\frac{\sigma u}{\pi d}\hat{z}.
\end{align}

\section{Hill equation and stability}\label{AppendixD}
We can now insert Eq.~\eqref{eq:force} into Eq.~\eqref{eq:final} to find the
governing equation of a single mode as
\begin{align}
  \partial_{tt}p=-\omega^2_n\left[1+\left(\frac{\mu_0q}{4}\right)^2\frac{\sigma}{\pi Mg}\frac{ A\omega\cos(\omega t)}{d_0+A\sin(\omega t)}\right]p,
\end{align}
where we have assumed that the conducting plate position is given by $z(t)=A\sin(\omega t)$ and its
motion is larger than the magnet one, much like the linearized equations of motion found in Ref.~\cite{Boeck2020}.

We introduce now $\tau=\omega t$ and perform a change of variables
\begin{align}\label{eq:11}
  \partial_{\tau\tau}p=-\left[\alpha+\gamma f(\tau)\right]p,
\end{align}
where
\begin{align}
  \alpha&=\frac{\omega^2_n}{\omega^2},  \quad \gamma = \left(\frac{\mu_0q}{4}\right)^2\frac{\sigma\omega_n^2}{\pi\omega Mg} \\
  f(\tau)&=\frac{\beta\cos(\tau)}{1+\beta\sin(\tau)}, \quad \beta=\frac{A}{d_0}.
\end{align}
In order to analytically compute the approximative instability regions of Eq.~\eqref{eq:11}, we expand the
function $f(\tau)$ into a Fourier series up to the second order. This yields
\begin{align}
  \partial_{\tau\tau}p=-\left[\alpha+\gamma\left(e_1\cos \tau+e_2\sin 2\tau\right)\right]p-\kappa\partial_\tau p,
\end{align}
where a damping term $\kappa$ is now included. Both $e_1$ and $e_2$ are functions of $\beta$
and are given in Eq.~\eqref{eq:Coeff}. We consider $p=\sum_{-\infty}^\infty a_n e^{in\tau/2}$, giving us a system of
equations for $a_n$. We truncate the system to two dimensions and impose that the corresponding
determinant is zero, i.e.,
\begin{align}
  \begin{vmatrix}
\alpha-\frac{1}{4}+\frac{i}{2}\kappa & \frac{1}{2}\gamma e_1 \\
\frac{1}{2}\gamma e_1 & \alpha-\frac{1}{4}-\frac{i}{2}\kappa
\end{vmatrix}
\end{align}
giving
\begin{align}
  \left(\alpha-\frac{1}{4}\right)^2+\frac{1}{4}\kappa^2-\frac{1}{4}(\gamma e_1)^2=0.
\end{align}
We repeat the procedure for the harmonic solutions with $p=\sum_{-\infty}^\infty a_n e^{in\tau}$, except that a larger
dimension is needed now
\begin{align}
  \begin{vmatrix}
2(\alpha-1+i\kappa) & \gamma e_1 & -i\gamma e_2 \\
\gamma e_1 & 2\alpha & \gamma e_1 \\
i\gamma e_2 & \gamma e_1 & 2(\alpha-1-i\kappa)
\end{vmatrix}
\end{align}
which leads to
\begin{align}\label{eq:solharmo}
  \alpha\left[4(\alpha-1)^2+4\kappa^2-\gamma^2e_2^2\right]-2(\alpha-1)\gamma^2e_1^2=0.
\end{align}

\section{Fourier coefficients}\label{AppendixE}
The function $f(\tau)$ is expanded as
\begin{align}
  f(\tau)\approx e_1\cos(\tau)+e_2\sin(2\tau)
\end{align}
with
\begin{equation}
\begin{split}\label{eq:Coeff}
  e_2 =&  \frac{2}{\beta^2} \left(\beta^2+2 \sqrt{1-\beta^2}-2\right) \\
  e_1 =& \frac{r}{2\pi \beta\sqrt{1-\beta}} \\
  r =& 4 \pi  \left(2 \sqrt{\beta+1} \beta+\sqrt{1-\beta} -2 \sqrt{\beta+1}\right)  \\ 
  & -8 (\beta-1) \sqrt{\beta+1} \tan ^{-1}\left(\sqrt{\frac{1-\beta}{\beta+1}}\right)   \\
   & -8 (\beta-1) \sqrt{\beta+1} \tan^{-1}\left(\sqrt{\frac{\beta+1}{1-\beta}}\right)
\end{split}
\end{equation}

\section*{Declaration of competing interest}
The authors declare that they have no known competing financial interests or personal relationships that could have appeared to influence the work reported in this paper.

\section*{Data availability}
Data will be made available upon request


\begin{thebibliography}{100}

\bibitem{Bernoulli1738}D. Bernoulli, Theoremata de oscillationibus corporum filo flexili connexorum et catenae verticaliter suspensae, Comm. Acad. Scient. Petrop. {\bf 6}, 108 (1738); L. Euler, De oscillationibus fili flexilis quotcunque pondusculis onusti, Comm. Acad. Sci. Petrop. {\bf 8}, 30 (1741); See J.T. Cannon and S. Dostrovsky,  (1981). ``Daniel Bernoulli (1733; 1734); Euler (1736)'' \href{https://doi.org/10.1007/978-1-4613-9461-7_9}{In: The Evolution of Dynamics: Vibration Theory from 1687 to 1742, Ch. 9, pp. 53 - 69  (Springer, New York, NY 1981)}. 

\bibitem{ArmstrongAJP1976}H. L. Armstrong, Effect of the mass of the cord on the period of a simple pendulum, \href{http://dx.doi.org/10.1119/1.10378}{Am. J. Phys. {\bf 44}, 564 (1976)}; C. G. Montgomery, Pendulum on a massive cord, \href{http://dx.doi.org/10.1119/1.11335}{Am. J. Phys. {\bf46}, 411 (1978)}; R. I. Sujith and D. H. Hodges, Exact solution for the free vibration of a hanging cord with a tip mass, \href{http://dx.doi.org/10.1006/jsvi.1995.0022}{J. Sound Vib. {\bf 179}, 359 (1995)}
\bibitem{Deschaine2008}J.~S. Deschaine and B.~H. Suits, The hanging cord with a real tip mass, \href{https://doi.org/10.1088/0143-0807/29/6/010}{Eur. J. Phys. {\bf 29}, 1211 (2008)}.
\bibitem{ColtmanJASA1979}J. W. Coltman, Acoustical analysis of the Boehm flute, \href{https://doi.org/10.1121/1.382350}{J. Acoust. Soc. Am. {\bf 65}, 499 (1979)}. 
\bibitem{LiuPRL1992}C.-H. Liu and S. R. Nagel, Sound in Sand, \href{https://doi.org/10.1103/PhysRevLett.68.2301}{Phys. Rev. Lett.  {\bf 68}, 2301 (1992)}.

\bibitem{Lee1970}S. M. Lee, The Double-Simple Pendulum Problem, \href{https://doi.org/10.1119/1.1976384}{Am. J. Phys. {\bf 38}, 536 (1970)}; R. B. Levien and S. M. Tan, Double pendulum: An experiment in chaos, \href{https://doi.org/10.1119/1.17335}{Am. J. Phys. {\bf 61}, 1038 (1993)}.
M. Wojna, A. Wijata, G. Wasilewski, and J. Awrejcewicz, Numerical and experimental study of a double physical pendulum with magnetic interaction,
 \href{https://doi.org/10.1016/j.jsv.2018.05.032}{J. Sound Vib. {\bf 430}, 214 (2018)}.

\bibitem{Boeck2020}T. Boeck, S.~L. Sanjari, and T. Becker, Parametric instability of a magnetic pendulum in the presence of a vibrating conducting plate, \href{https://doi.org/10.1007/s11071-020-06054-y}{Nonlinear Dyn. {\bf 102}, 2039 (2020)}.

\bibitem{Schmidt1984}V.~H. Schmidt and B.~R. Childers, Magnetic pendulum apparatus for analog demonstration of first‐order and second‐order phase transitions and tricritical points, \href{https://doi.org/10.1119/1.13847}{Am. J. Phys. {\bf 52}, 39 (1984)}.
\bibitem{Kim1997}S.-Y. Kim, S.-H. Shin, J. Yi, and C.-W. Jang, Bifurcations in a parametrically forced magnetic pendulum, \href{https://doi.org/10.1103/PhysRevE.56.6613}{Phys. Rev. E {\bf 56}, 6613 (1997)}.
\bibitem{Starrett1974}J. Starrett and R. Tagg, Control of a chaotic parametrically driven pendulum, \href{https://doi.org/10.1103/PhysRevLett.74.1974}{Phys. Rev. Lett. {\bf 74}, 1974 (1995)}.
\bibitem{Luo2020}Y. Luo, W. Fan, C. Feng, S. Wang, and Y.Wang, Subharmonic frequency response in a magnetic pendulum, \href{https://doi.org/10.1119/10.0000038}{Am. J. Phys. {\bf 88}, 115 (2020)}.
\bibitem{Khomeriki2016}G. Khomeriki, Parametric resonance induced chaos in magnetic damped driven pendulum, \href{https://doi.org/10.1016/j.physleta.2016.05.049}{Phys. Lett. A {\bf 380}, 2382 (2016)}.
\bibitem{Shliomis2004}M.~I. Shliomis and M.~A. Zaks, Nonlinear dynamics of a ferrofluid pendulum, \href{https://doi.org/10.1103/PhysRevLett.93.047202}{Phys. Rev. Lett. {\bf 93} 047202 (2004)}.

\bibitem{Belmonte2001}A. Belmonte, M.~J. Shelley, S.~T. Eldakar, and C.~H. Wiggins, Dynamic patterns and self-knotting of a driven hanging chain, \href{https://doi.org/10.1103/PhysRevLett.87.114301}{Phys. Rev. Lett. {\bf 87}, 114301 (2001)}.


\bibitem{Landau1960}L.~D. Landau and E.~M. Lifshitz, \textit{Mechanics} (Pergamon Press, 1960).

 \bibitem{Abou1992}A. M. Abou-Rayan, A. H. Nayfeh, D. T. Mook, M. A. Nayfeh, Nonlinear response of a parametrically excited buckled beam, \href{https://doi.org/10.1007/BF00053693}{Nonlinear Dyn.  {\bf 4}, 499 (1993)}.

  
\bibitem{Jackson}J. D. Jackson, {\em Classical Electrodynamics}, 3rd ed. (J. Wiley \& Sons Inc., New York, 1999).


\bibitem{Furlani001}E. P. Furlani, \textit{Permanent Magnet and Electromechanical Devices} (Academic Press, San Diego, 2001).

 \bibitem{Concha2018}A. Concha, D. Aguayo, P. Mellado, Designing Hysteresis with Dipolar Chains, \href{https://link.aps.org/doi/10.1103/PhysRevLett.120.157202}{Phys. Rev. Lett.  {\bf 120}, 157202 (2018)}.
  
\bibitem{Chicone2006}C.~Chicone, \textit{Ordinary Differential Equations with Applications, Texts in Applied Mathematics} (Springer, New York, 2006).

\bibitem{Verhulst2022}F. Verhulst, Perturbation Analysis of Parametric Resonance, \href{https://doi.org/10.1007/978-1-0716-2621-4_393}{In: Perturbation Theory: Mathematics, Methods and Applications, pp. 167 - 183 (Springer, New York, NY 2022).}


\end{thebibliography}

\end{document}